\documentclass[11pt,twoside]{article}

\usepackage{asp2006}
\usepackage{epsf}
\usepackage{lscape}

\def\deg{$^{\circ}$~}
\def\arcsec{$\,^{\prime\prime}$~}

\def\Msun{$M_\odot$}
\newcommand{\lsim }{{\lower0.8ex\hbox{$\buildrel <\over\sim$}}}
\newcommand{\gsim }{{\lower0.8ex\hbox{$\buildrel >\over\sim$}}}

\begin{document}
\title{AGN Variability Surveys: DASCH from BATSS to EXIST}   
\author{Jonathan E. Grindlay}   
\affil{Harvard-Smithsonian Center for Astrophysics, 60 Garden St.,
  Cambridge, MA 02138, USA} 

\begin{abstract} %%% Abstract to run on from here.
Active galactic nuclei (AGN) are variable on a wide range 
of timescales, though relatively few systematic variability surveys 
have been conducted. Previous broad-band (both 
spectral and temporal) variability surveys of AGN 
are limited in their temporal and spectral bandwidth, 
despite their promise for probing the central 
engine and black hole mass. We outline optimal properties for 
variability studies and provide a brief 
summary of three new variability 
surveys, two (BATSS and DASCH) about 
to begin and the third (EXIST) possible within the next decade, which 
will open  new windows on the physics and fundamental properties of AGN.
\end{abstract}

%%% MAIN BODY OF TEXT GOES HERE. CONSULT "INSTRUCTIONS FOR AUTHORS USING
%%% LATEX2E MARKUP", SECTIONS 2.3-2.6 FOR HELP WITH EQUATIONS, FIGURES,
%%% AND TABLES.

\section{Introduction}   
A number of AGN studies have been undertaken for the variability 
of both optical (e.g. Peterson, these proceedings) 
and X-ray fluxes (e.g. Uttley, these proceedings) 
to enable constraints on the mass of the central supermassive 
black hole (SMBH) and to constrain emission models 
for the central engine, accretion disk and jet(s). Most 
of these have been with targeted narrow-field telescopes with 
observational cadence and total duration necessarily limited. The 
ideal AGN variability survey (AVS) would have the following 
properties, Pn: 
\begin{enumerate}
\item {\it AVS-P1:} {\it broad sky coverage}, $\Omega$, to maximize the 
number of AGN observed and to enable rare classes of variable 
objects, and low duty cycle events, to be found; 
\item {\it AVS-P2:} {\it long total survey duration}, {\it D}, 
of survey observations which each detect minimum source flux 
S$_{min}$ on timescale $\tau_o$ and with fractional uncertainty 
in flux $\delta S_{\tau_o}$ and 
enable variations on timescales from a maximum  
$\tau_{max} \sim${\it  D/2}  
down to a minimum (median) $\tau_{min} \sim0.7D/N$ to be measured 
from N randomly sampled observations; and 
\item {\it AVS-P3:} {\it large total number of measurements}, 
{\it N}, to enable measures of source variability 
on timescales $\tau_n \sim (n - 1)\tau_{min}/2$, where n = 2 ...N, 
and fractional variability sensitivity improves as 
 $\delta S_{\tau_n} \sim  \delta S_{\tau_o}/n^{0.5}$
\end{enumerate}
 
Previous AGN variability studies have typically met only AVS-P3, and 
then usually with relatively short duration D. Broad-field (AVS-P1) AGN 
variability surveys are almost unknown, though the Swift/BAT 
survey (Markwardt et al 2005) with $\sim$70\% sky coverage 
per day has begun to open up this domain.
Here we outline three new surveys that will each extend one or more 
of these AVS properties.

\section{New AGN variability surveys}
Time domain studies (e.g. PanStarrs and LSST) will unleash  new  
constraints on AGN parameters and models by having  
one or more of the broad properties listed above. Already, optical 
and hard X-ray timing studies for full-sky AGN samples 
are beginning, and a far-reaching 
X-ray/$\gamma$-ray temporal-spectral survey could {\it EXIST} as an 
Einstein Probe in NASA's Beyond Einstein Program.

\subsection{DASCH: Optical variability on scales $\Delta\tau \sim$ 10d
--(50-100y)} Over the past 3y, we have developed an astronomical plate digitizer 
(Simcoe et al 2006) that is some 100X faster than any previously built
in order to make available the Digital Access to a Sky Century from 
Harvard ({\it DASCH}). This will make possible (grant funding or a donor 
permitting...) the digitization and on-line access to the full 
images and derived photometry of Harvard's unique collection of 
some 600,000 astronomical images (all at least 5\deg x 7\deg) of the 
full northern and southern sky from c. 1880 - 1985. Astrometry to 
\lsim1\arcsec is derived from WCS solutions for each scan and 
photometry from SExtractor isophotal analysis fits calibrated locally 
from the \gsim3000 GSC2.2 stars (B $\sim$8-15) typically on each 
plate (Laycock et al 2007). An example light curve 
(Fig. 1) for a random star in the open cluster M44, 
used for development of photometric analysis software, shows 
the \lsim0.1mag (rms) photometry possible which 
for this dataset from 5 different plate series over 88y.  
\begin{figure}[!h]
\plotone{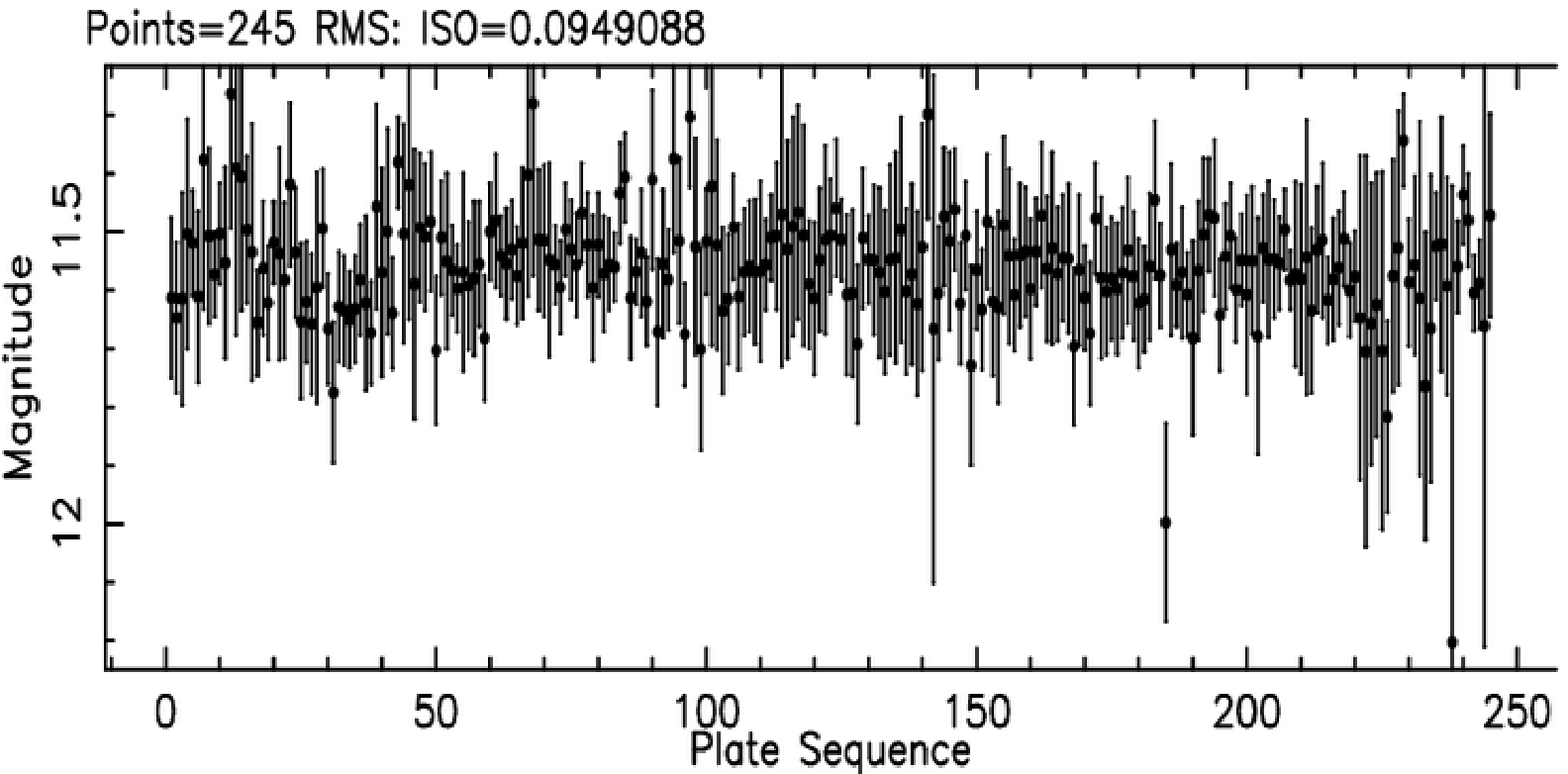}
\begin{caption}
{Partial lighturve (B mag vs. plate no.) for star (\#399=GSC1395-2447) 
in the Open Cluster M44 measured on 245 plates (1890 - 1978). Even 
smaller rms errors are likely when annular calibration (on GSC2.2) is 
used vs. the full plate average done here since psf variations are 
then included.}
\end{caption}
\end{figure}
Analysis of an initial sample of 15 bright (B \lsim15) PG QSOs, 
starting with 3C273, is planned for a pilot study of variability power 
density spectra (PDS) to explore PDS break 
timescales $\tau_{bk}$ as a measure 
of SMBH mass. With $\sim$1000 plates for any given object 
randomly observed over $\sim$100y, the median sampling time is 
$\sim$25d and so the possible variability timescale range is 
$\tau_{max}/\tau_{min} \sim 50y/25d \sim730$. Allowing for \gsim3 
timescale measures above a PDS break to determine $\tau_{bk}$, the 
corresponding SMBH mass range could be constrained over 
dynamic range of $\sim$240.  

\subsection{BATSS: Hard X-ray variability on scales $\tau 
 ~{\hbox{\rlap{\raise.3ex\hbox{$<$}}\lower.8ex\hbox{$\sim$}\ }}$
100s--1d} We have also initiated a ``BAT Slew Survey'', 
BATSS (Copete et al 2007) using the BAT hard 
X-ray imager (Barthelmy et al 2005) on Swift 
to analyze ``event-mode'' data from the 
$\sim$60 slews ($\sim$1-2min each) that Swift performs each 
day to slew on/off pointed targets. Whereas BAT pointings cover 
some $\sim$70\% of the sky each day, adding in the slews increases 
sky coverage to nearly 100\% as well as provides the only high 
time resolution data (apart from GRBs) since BAT pointing data 
is binned on 5-7min timescales. BATSS will thus provide AVS-P1,P3 
and be particularly well suited to detect rare, bright AGN flares 
such as the extreme Blazar events from PKS2155-304 for which 
Swift/XRT/BAT coverage did not quite overlap 
with the  ~8 and ~17Crab(!) TeV flares reported by 
HESS (Foschini et al 2007). Although 
the XRT spectra indicate that the synchrotron 
spectral break for this Blazar is below the BAT band, 
the BAT Transient Monitor (Krimm 2007) clearly does 
see flare variability from others -- e.g. Blazar Mrk 421 (Fig. 2) 
for which extreme flares could be seen by BATSS.  

\begin{figure}[!h]
\plotfiddle{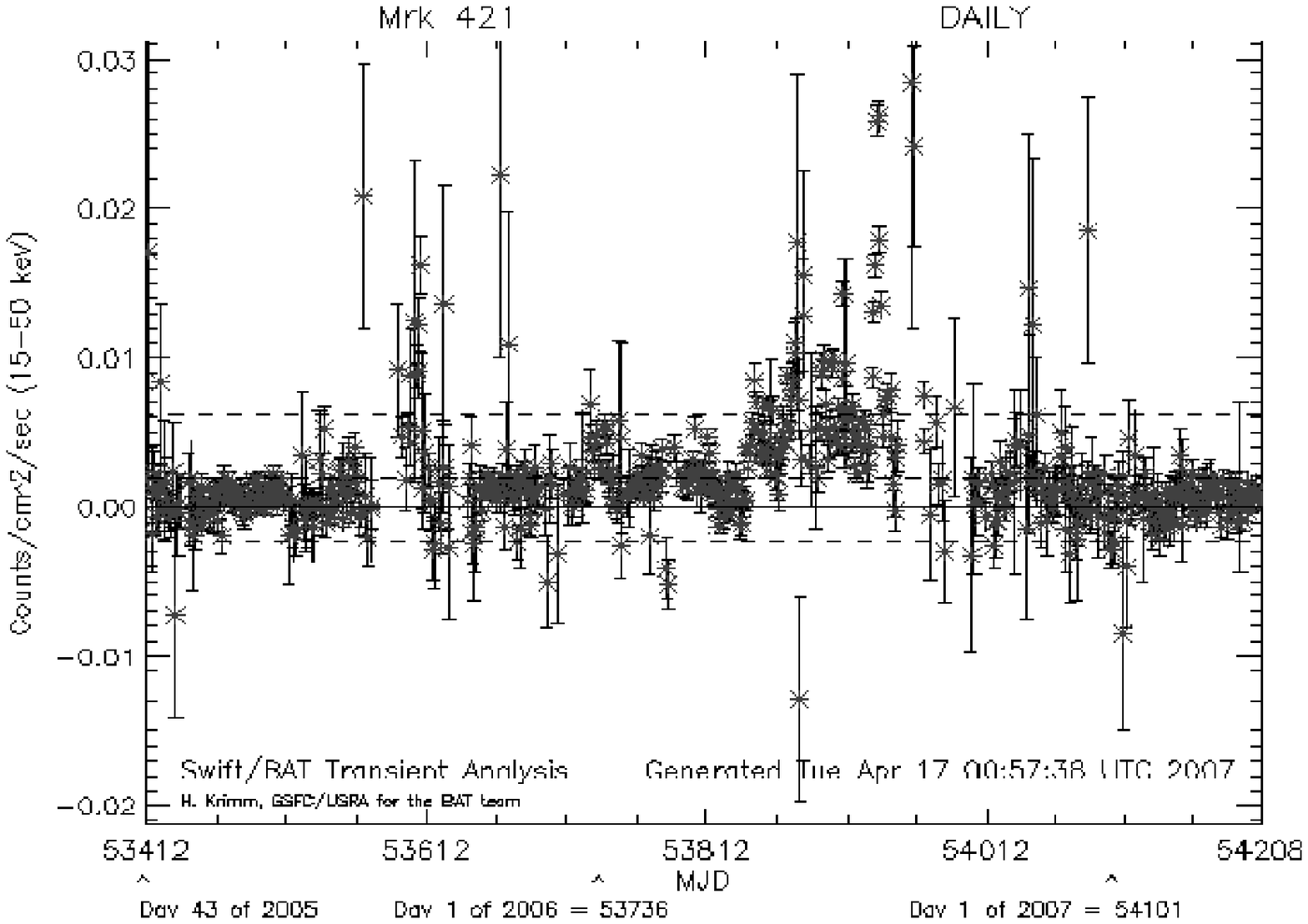}{2.8in}{0.}{70.}{70.}{-170.}{-10.}
%\plotone{mrk421-BAT-lc.eps}
\begin{caption}
{Swift/BAT 2y lightcurve for Blazar Mrk421 showing $\sim$100mCrab 
flares that might be resolvable (and still brighter) by BATSS.} 
\end{caption}
\end{figure}

\subsection{EXIST: Ultimate hard X-ray variability on scales $\tau
 ~{\hbox{\rlap{\raise.3ex\hbox{$<$}}\lower.8ex\hbox{$\sim$}\ }}$
10s--5y}  The best prospects to optimize properties AVS-P1-P3 are 
with the {\it EXIST} mission (http://exist.gsfc.nasa.gov/), proposed 
as the Black Hole Finder Probe in NASA's Beyond Einstein Program. 
{\it EXIST} images the full sky 3-600 keV each 95min orbit with two 
large area and field of view (FoV) coded aperture telescopes 
(Grindlay 2005 and Grindlay et al 2007). With  
daily full-sky flux sensitivity $S_{min} \sim$1mCrab  
(comparable to Swift/BAT in 1y) due to nearly 20\% continuous 
coverage on every source enabled by continuous scanning with  
its large FoV and total area, {\it EXIST} would detect and 
study \gsim3 x 10$^4$ AGN full sky. Each is located to 
\lsim 11\arcsec (90\% confidence radius) 
which allows unambiguous host galaxy 
identification for its 0.05mCrab (=5 x 10$^{-13}$ cgs, 40-80 keV) 
5$\sigma$ survey threshold sources. A simulated 1y survey 
image and logN-logS is shown in Fig. 3 with normalization 
from Treister and Urry (2005) including obscured AGN. Thus, 
$\sim$300 AGN (full sky) can be measured on timescales 
$\tau_{min}$ = 1d or $\sim$1000 AGN with $\tau_{min}$ = 6d. For 
a 5y mission, AVS-P3 gives N = 1800 and 300 timescales, respectively,  
to constrain the PDS and $\tau_{bk}$ and thus SMBH mass. 
\begin{figure}
\plottwo{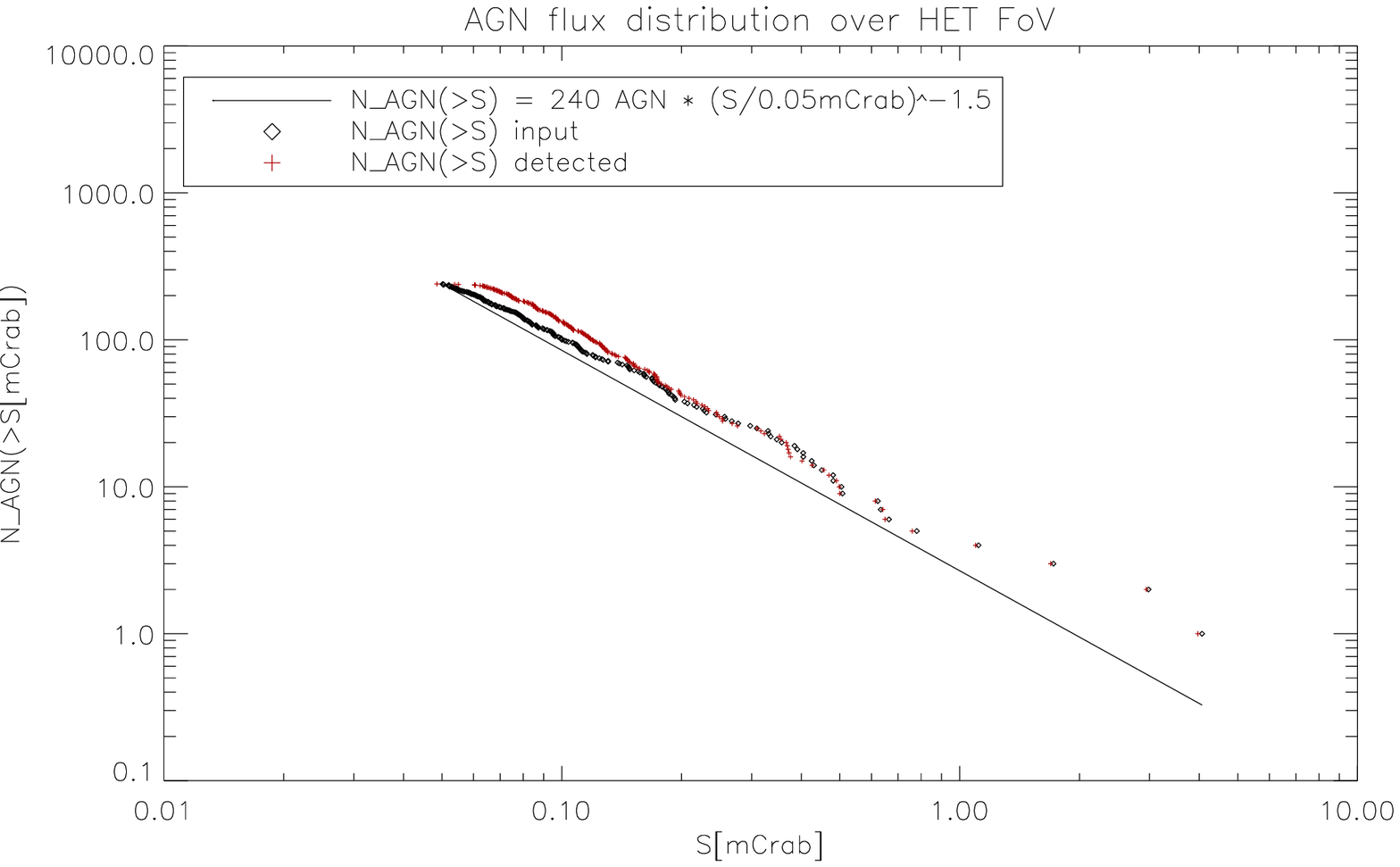}{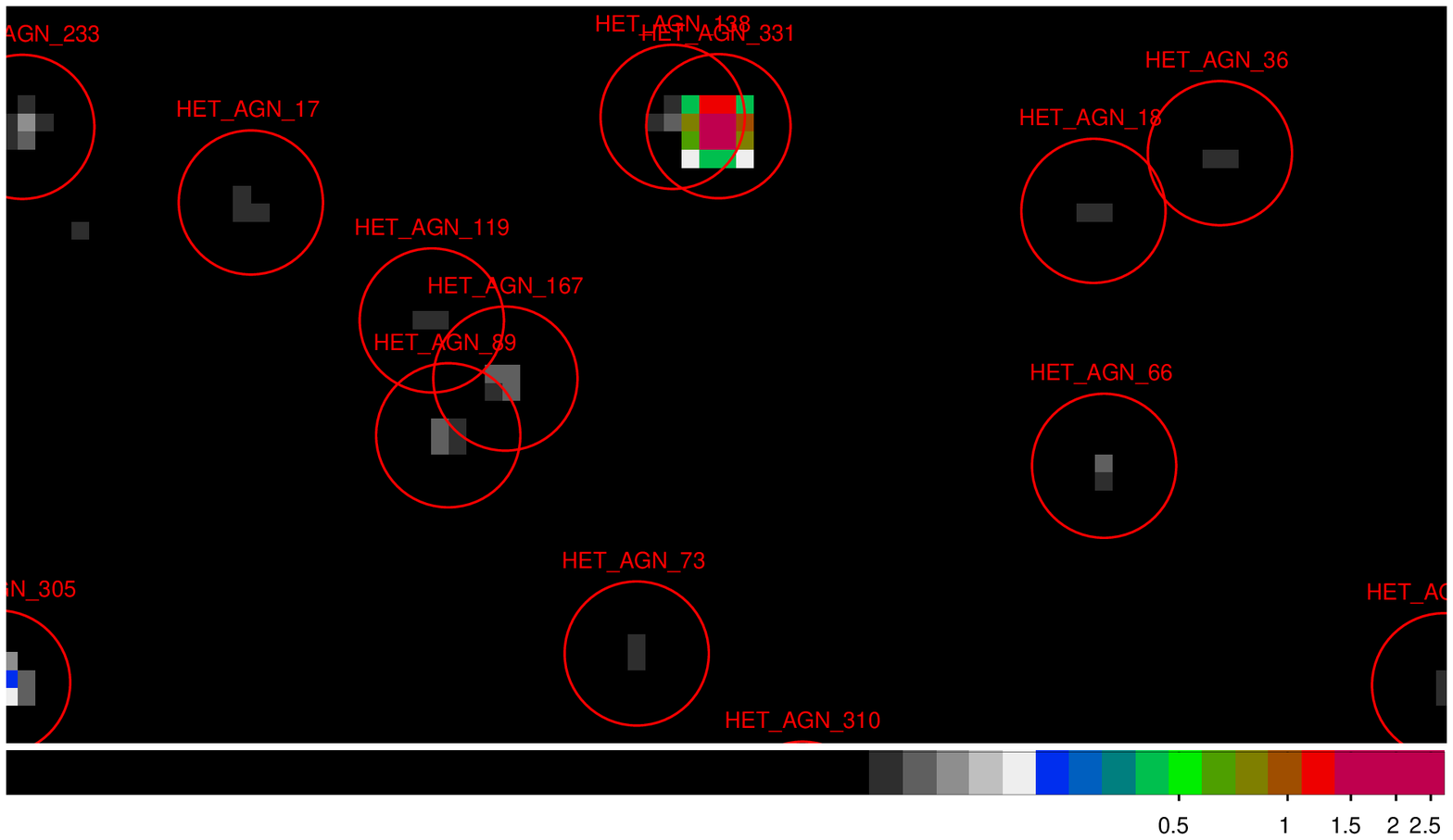}
\begin{caption}
{Simulated 1y {\it EXIST} logN-logS for 21\deg x 21\deg field of 
view of one sub-telescope containing 331 AGN and 4\deg x 2\deg zoom 
image including brightest (4mCrab) 
source and 12 other AGN with fluxes from $\sim$0.05 -- 0.28mCrab.}  
\end{caption}
\end{figure}
 
\vspace*{-0.3cm}

\section{Prospects for SMBH mass, demographics and evolution}
Given the $\tau_{bk}$ vs. mass scaling found by Uttley et al (2002) 
from RXTE data on several AGN, {\it EXIST} and possibly DASCH can 
constrain SMBH masses in AGN over a range $\sim$10$^{7-9.5}$ \Msun,   
with \gsim10$^{9}$\Msun traced out to z \lsim3. 
Non-AGN SMBHs (e.g. SgrA*) can be identified by their 
tidal disruption of main sequence stars and accompanying 
hard X-ray flares detected by {\it EXIST} 
and optical flares detected by DASCH for non-obscured systems. 
Together with extreme Blazar events from BATSS, the prospects 
for new AGN variability surveys are timely.

{}

%\section*{}    %%% Unnumbered top level section head (remove "%" symbol)
%\subsection*{}   %%% Unnumbered second level section head (remove "%" symbol)

\vspace*{-0.3cm}

\acknowledgements %%% Text of acknowledgements runs on after this command.
I thank S. Laycock and S. Tang and our DASCH Team (NSF grant 
AST-0407380), A. Copete and our 
BATSS Team (NASA grant NNX07AF78G) 
and J. Hong and the EXIST Team  
(NASA grant NNG04GK33G). 

%%% THE BIBLIOGRAPHY
%%%
%%% CONSULT SECTION 3 OF "INSTRUCTIONS FOR AUTHORS" FOR HOW TO USE NATBIB.
%%% AUTHORS ARE ENCOURAGED TO USE EITHER THE "THEBIBLIOGRAPY" ENVIRONMENT
%%% BY UNCOMMENTING (DELETING THE "%" SYMBOL) THE COMMANDS BELOW, OR BY
%%% USING THE BIBTEX ENVIRONMENT. TO FIND OUT WHICH IS APPLICABLE TO YOUR
%%% CONTRIBUTION, CONSULT THE VOLUME EDITORS FOR YOUR PROCEEDINGS.
%%%

%\begin{thebibliography}{}
%\bibitem[]{}
%\bibitem[]{}
%\bibitem[]{}
%\bibitem[]{}
%\bibitem[]{}
%\bibitem[]{}
%\bibitem[]{}
%\bibitem[]{}
%\bibitem[]{}
%\bibitem[]{}
%\bibitem[]{}
%\bibitem[]{}
%\end{thebibliography}

\end{document}